\documentclass[a4paper]{jpconf}

\usepackage{amsfonts}
\usepackage{amsbsy} 
\usepackage{epsfig}
\usepackage{latexsym}
\usepackage{graphicx}
\usepackage{enumerate}

\input amssym.def
\input amssym.tex

\def\barr{\begin{array}}
\def\earr{\end{array}}

\def\be{\begin{equation}}
\def\ee{\end{equation}}
\def\bea{\begin{eqnarray}}
\def\eea{\end{eqnarray}}

\newcommand{\cG}{{\mathcal G}}
\newcommand{\cH}{{\mathcal H}}

\newcommand{\cJ}{{\mathcal J}}

\newcommand\N{{\mathbb N}}
\newcommand\R{{\mathbb R}}

\newcommand{\su}{\mathfrak{su}}
\newcommand{\SU}{{\rm SU}}

\def\extd{\mathrm {d}}

\def\vphihat{\widehat{\varphi}}

\newcommand\acts\triangleright

\begin{document}

\title{Ten questions on Group Field Theory \\ 
(and their tentative answers)}
\author{Aristide Baratin, Daniele Oriti}
\ead{aristide.baratin@aei.mpg.de, daniele.oriti@aei.mpg.de}
\address{Max Planck Institute for Gravitational Physics (Albert Einstein Institute), Am M\"uhlenberg 1, D-14476 Golm, Germany, EU}

\begin{abstract}
We provide a short and non-technical summary of our current knowledge and some possible perspectives on the group field theory formalism for quantum gravity, in the form of a (partial) FAQ (with answers). Some of the questions and answers relate to aspects of the formalism that concern loop quantum gravity. 
This summary also aims at giving a brief, rough guide to the recent literature on group field theory (and tensor models).
\end{abstract}


\section{Introduction}

\

\noindent The aim of this paper is to provide a short summary of what we know about the group field theory (GFT) formalism for quantum gravity and  its longer term goals. For greater agility of presentation and reading, the format is that of a FAQ: we outline tentative and provisional answers to some basic questions about the GFT formalism, which  aim at reflecting the current understanding of the subject, 
from the strict point of view of the authors. 

Having clarified our aims, let us clarify what this paper is {\sl not} meant to be.  
First, it is not meant to be a technical paper dwelling into any of the details of the formalism, or explaining in detail any of the recent results. 
For this we will refer to the relevant literature. 
Second, it is not a pedagogical, extensive introduction to GFTs. 
For early introductions,  the reader should go to \cite{laurentgft, iogft}, and for the general idea of tensor models 
one should look at the early papers  \cite{tensor} and at the literature on matrix models \cite{mm}. 
For a more extensive, and almost up to date introduction to GFTs for quantum gravity, we refer to \cite{ProcCapeTown}.
Third, it is not a proper review of the subject nor of its recent developments, which are many and important. 
For many of the recent results on (colored) tensor models, a  complete and up-to-date (and beautifully written) review is \cite{TensorReview}. 
For many of the recent results that relate to the geometry of group field theories,  we refer to the review \cite{ioaristideReview}.
For a discussion of spin foam models, in particular the recent ones, and thus some important aspects of the corresponding group field theories, we refer to the many reviews on spin foams \cite{SF}. 
Many other results, concerning for example GFT and tensor model renormalization 
can unfortunately only be found in the original papers.

\section{Ten questions, with their tentative answers}

\

(i) {\bf  What is a group field theory?}

\

\noindent Group field theories are a specific class of tensor quantum field theories which generalize matrix models to higher dimensions. 
The fundamental variables are in fact higher rank {\it tensors} formally denoted by $T_{i_1 ... i_d}$ where the indices label points in the direct product of d domain spaces. 
The domain space could be a manifold or a finite set. The dynamics of tensor models is governed by a classical action characterized by a peculiar pairing of the indices in the interaction term:
\[
S_{int}(T) = \lambda \sum_{\vec{i}^{j}} \, T_{\vec{i}^1} \cdots T_{\vec{i}^{d+1}} \vspace{-0.2cm}
\]
where $\vec{i}^j\!=\!(i^{jj-1} ,..., i^{j0}, i^{jd} ,..., i^{jj+1})$ is the set of indices of the $j$ copy of the tensor in the monomial of degree $d+1$, with $i^{jk}=i^{kj}$.
For example the action for the independent identically distributed 3-tensor model is 
\be \label{tensor} S(T) = \frac12 \sum_{i,j,k} T_{ijk} T_{kji} \, - \, \frac{\lambda}{4!}\, \sum_{ijklmn} T_{ijk} T_{klm} T_{mjn} T_{nli}\;  \ee
A tensor can be graphically represented as a d-valent node with an index on each line, or dually as a (d-1)-simplex with an index on  each (d-2)-face. 
The interaction term pairs the indices along the links of a network with d+1 d-valent nodes; dually  it patterns the glueing of d+1 (d-1)-simplices along common faces to form a d-simplex. The kinetric term describes instead the identification of two nodes or (d-1)-simplices. 
Different choices of kinetic and interaction terms, 
are possible and define different models in the same class. 
The peculiar interaction combinatorics of any tensor model, shows up in the perturbative expansion of the theory in Feynman diagrams
$$ Z\,=\,\int
\mathcal{D} T\,e^{-S[T]}\,=\,\sum_{\Gamma}\,\frac{\lambda^V}{sym[\Gamma]}\,Z(\Gamma),
$$
where $V$ is the number of interaction vertices in the Feynman graph
$\Gamma$, $sym[\Gamma]$ is a symmetry factor for the graph and
$Z(\Gamma)$ the corresponding Feynman amplitude. By construction the Feynman diagrams will be represented as stranded diagrams dual to cellular (simplicial) complexes of arbitrary topology, obtained by arbitrary glueings of d-simplices along (d-1)-faces. 

Tensors models are thus a straightforward combinatorial generalization of matrix models for 2d quantum gravity. 
In matrix models, the Feynman amplitudes associated to each diagram can be re-interpreted in terms of a discretization of 2d general relativity on an equilateral triangulation, and the overall sum as a generalization of a lattice path integral for 2d gravity to arbitrary triangulations and arbitrary topologies. 
Among the key results in matrix models \cite{mm} are the large N topological expansion dominated by trivial topologies (where N is the dimension of the matrix) and the identification of critical behaviour for some value of the coupling constant. This critical behaviour 
defines the continuum limit, shown by various methods to match the effective continuum dynamics to 2d Liouville quantum gravity. 
Tensors models try to export these ideas and results to higher dimensions. 
However, while even simple tensor models of the type (\ref{tensor}) have an interest for quantum gravity, as they relate directly to the approach of dynamical triangulations,  both the richness of spacetime geometry in higher dimensions (especially in 4 dimensions) and the results obtained by other approaches to quantum gravity suggest that a richer set of data and symmetries should be added to the tensors, and that a more involved dynamics should be chosen, to have a better chance to describe quantum geometry and gravity. 

Group field theories (GFT)  is the name used, in fact, for those tensor models where the domain space is chosen to be the local gauge group $G$ of gravity (i.e. the Lorentz group or its euclidean counterpart), so that  tensors thus turn into fields $\varphi \!\in\! L^2(G^d)$. An additional symmetry is invoked which captures local gauge invariance:
$$
\varphi(h g_1, \cdots h g_d) = \varphi(g_1 \cdots g_d) \quad \forall h\in G,
$$ 
and, possibly, kinetic and interaction terms are characterized by non-trivial kernels. 
In turn, alternative representations in terms of group representations or as non-commutative quantum field theories over $\mathbb{R}^{|G|}$, 
where $|G|$ is the dimension of $G$,  become available, thanks to the group structure and tools from representation theory and non-commutative geometry. 
An example is the Boulatov model for 3d Riemanian quantum gravity \cite{Boulatov}, where $G \!=\! \SU(2)$ and with action:
\be 
S_{3d}[\varphi]\,=\, \frac12 \int[\extd g]^3 \, \varphi_{123}\varphi_{321} - \, \frac{\lambda}{4!}\,\int [\extd g]^6 \,
\varphi_{123}\varphi_{345}\varphi_{526}\varphi_{641} \quad  \label{boulatov} 
\ee
where $\varphi_{ijk}$ is a short notation for $\varphi(g_i, g_j, g_k)$ and $\extd g$ is the Haar measure on the gauge group. 

The GFT formalism thus attempts to {\it define} a theory of quantum spacetime as a superposition of discrete spaces, each generated as a possible interaction process of fundamental building blocks, tentative quanta of space, by incorporating into the tensor model framework lessons and insights of other approaches to quantum geometry -- quantum Regge calculus, quantum simplicial geometry and Loop Quantum Gravity (LQG) \cite{lqg}, as well as mathematical tools taken from non-commutative geometry:
\begin{itemize} 
\item 
The insight from quantum simplicial geometry suggests the variables needed to describe a `quantum geometric (d-1)-simplex'  
which has to become the fundamental atom of quantum space, the basic `quantum' in our enriched tensor model. 
\item A characterization of the geometry and of the whole kinematical phase space to be associated to a simplex is also provided by the analysis of discretized gravity actions in various dimensions and in various formulations. In particular, 1st order formulations of gravity in terms of connection and (conjugate) vielbein variables suggest to use group and Lie algebra manifolds to define the phase space of such simplices, and the domain spaces of tensor models, leading then to group field theories. 
\item Loop Quantum Gravity provides further justification for the kinematical structures used for defining the Hilbert space of quantum states of these models, several insights on their interpretation, and a main source of inspiration for constructing models and specifying the correct dynamics.
\item Simplicial gravity path integrals, and in particular quantum Regge calculus, provide a template to analyze the corresponding Feynman amplitudes, and to evaluate to what extent they capture correctly the dynamics of discrete geometry that can be associated to each Feynman diagram/simplicial complex.
\item Tools from non-commutative geometry then provide the means to define different representations of the same models, and to analyze their properties. 
\end{itemize}

\

(ii) {\bf How does the GFT formalism relate to loop quantum gravity?}

\

\noindent 
Polynomial gauge invariant observables of the group  field theory correspond to  functions of a finite number of gauge group elements associated to 
the links of a (d-valent) graph (plus possible additional variables) satisfying a gauge invariance condition at each  node. 
A basis of such observables is written in terms of so-called spin networks functionals, 
generically labelled by a graph $(\Gamma, j_e, \imath_v)$ with edges $e$ and vertices $v$ respectively group representations $j_e$ and invariant tensors $\imath_v$.
GFT n-point functions can thus  be expressed in terms of LQG-like states. 
The exact matching holds for some models, at least in representation space (and as far as the spectrum of some quantum geometric operators is concerned). 

Intuitively, GFTs provide a `second quantized'  framework for LQG: spin network vertices become the quanta of a GFT field, 
and the LQG wave function associated to an individual vertex is turned to a field operator that creates/annihilates spin network vertices from/to a `no-space'  vacuum, where no geometrical nor topological structure is present. 
It is difficult, however, to go beyond this intuitive picture and construct a proper Fock space of spin network vertices: to do so, one has to face the issue of choosing a statistics for them, which is both a very difficult and very exotic question that has never been considered in the LQG literature. 

As far as the quantum dynamics of LQG is concerned,  the main link lies in the structure of the GFT Feynman amplitudes $I_{\cG}$. 
These can indeed be written in terms of {\sl spin foam models} \cite{SF}, which show up in LQG as a natural arena to encode the Hamiltonian evolution of spin-networks as a sum over histories. A `spin foam'  $(\mathcal{J}, j_f, \imath_e)$  is a higher dimensional analogue of a spin-network (a `history' of spin-network evolution): it is  a system of branching surfaces $\mathcal{J}$ taking the form of a 2-complex, with polygonal faces labeled by group representations $j_f$ and edges labeled by intertwiners $\imath_e$.  To each GFT Feynman diagram $\cG$ corresponds a 2-complex $\mathcal{J}$: vertices and edges of $\cG$ corresponds to vertices and edges of $\cJ$ and 
the loops of strands on $\cG$ corresponds to the 2d faces of $\cJ$. One can show that the GFT amplitudes (in a specific representation) takes the generic form of `spin foam amplitudes': 
\be
I_{\cG} = \sum_{j_f, \imath_e} \prod_f A_f(j_f) \prod_e A_e(j_f, \imath_e) \prod_v A_v(j_f, \imath_e)
\ee
characterized by a choice of local amplitudes $A_f, A_e$ and $A_v$ assigned to the faces, edges and vertices of the 2-complex $\mathcal{J}$ defined by $\cG$. 
For example, the Feynman amplitudes of the Boulatov model (\ref{boulatov}) give the Ponzano-Regge model for 3d quantum gravity. 
This duality between GFT and spin foam amplitudes is general: the amplitudes of any spin foam model can be obtained as the Feynman amplitude of a GFT \cite{SF-GFT}. 

In a covariant, sum-over-histories  formulation of loop quantum gravity, both the combinatorial and algebraic data that specify a quantum state of geometry are randomized with a quantum probability amplitude in order to define the full quantum theory. GFTs thus provide a definition of this quantum dynamics, in the form of a quantum field theory expansion in possible interaction processes for the same building blocks of quantum geometry, which includes a sum over all topologies. 


The main object of interest for the LQG dynamics is the physical inner product between states, projecting onto the kernel of the Hamiltonian constraint. 
Indications of where this object should be found in the GFT formalism stem  from the interpretation of GFT as a second quantized formalism for spin network vertices. First, one would expect that some form of Hamiltonian constraint acting on spin network vertices can already be identified at the level of the GFT classical action and equations of motion.  
Second, as we mentioned, the dynamics encoded into a single evolution history of a spin network state can be found in all its details in the form of a GFT Feynman amplitude. 
Obviously, from the field theory perspective, this is but a tiny corner of the true quantum dynamics, which has to be looked for in the Schwinger-Dyson equations (SDE) for the n-point functions of the GFT model \cite{SDEOoguri, laurentgft}. In other words, one would expect that the SDE  for polynomial field correlations, that is spin network functionals, would admit (at least in a regime suppressing topology change) an interpretation and  a transcription as Hamiltonian and diffeomorphism constraints of a quantum gravity theory, and thus provide the definition of the physical inner product. 
This is indeed what happens, in the appropriate limit, in the simple case of matrix models. 
In particular, this should be true in the continuum limit of the GFT model, which is necessarily a non-perturbative domain. The same continuum dynamics can be also looked for at the level of effective actions in this continuum limit, at the level of symmetries satisfied by the model and, if a phase transition is part of the picture, in the nature of the critical exponents for the scaling of interesting observables.

 \

(iii) {\bf Why is the GFT formalism useful to LQG?}

\

\noindent This way of defining the quantum dynamics is advantageous, from a purely LQG perspective, for a variety of reasons. The main one is that a field theory framework is the most convenient one for dealing with the infinite number of degrees of freedom which 
should ultimately be present in the full continuum theory: a generic continuum geometry will in fact be captured by a highly complicated (superposition of) spin network state(s), that is, in GFT language, a hugely populated many-particle state. 
Furthermore, contrary to direct attempts at the definition of the continuum gravitational dynamics from the canonical quantum gravity perspective, the GFT formalism offers a compact, but in principle complete definition of the theory: it is given by the path integral of the GFT model. The difficulty is of course to go beyond the formal definition, prove its well-posedness, and extract interesting physics from it. 

The GFT definition also opens the door to the application of more or less standard quantum and statistical field theory tools (and ideas) to the study of the same dynamics, a very powerful machinery indeed. Last, it permits to LQG structures and results to force their way out of the 
\lq\lq canonical quantization of continuum gravity cage \rq\rq  , which is not only technically extremely challenging, but also suspicious for a variety of reasons. Several arguments indeed (very different in nature, each not conclusive, and still each very reasonable) can be put forward suggesting that general relativity (GR) is but an effective field theory, and that it should not be quantized 
as such, but only emerge in some corner of a fundamental theory defined in altogether different terms. GFTs allow in principle to test and, if realized, to describe in detail this scenario. 

\

(iv) {\bf How to define the right quantum dynamics of geometry as a GFT?}

\

\noindent Given the ingredients entering the very definition of a GFT model, the most natural way of proceeding to its construction relies mainly on quantum simplicial geometry. 
It amounts to answering three questions:
\begin{enumerate}[a)]
\item  what is a quantum geometric (d-1)-simplex? 
\item what is a quantum geometric d-simplex, or, equivalently, how to glue d+1 (d-1)-simplices to form a d-simplex? 
\item  how does the geometric information of one d-simplex propagate to a neighboring one? 
\end{enumerate}
The answer to a) dictates the nature  and properties of the GFT field and its variables. 
The answer to b) indicates what interaction kernel one should use to define the classical action of the corresponding GFT. 
The answer to c) gives the form of the kinetic term of the action, that is, the propagator of the field theory. 
Having all the ingredients at hand, one can write down the GFT action, and thus the full model, in perturbative expansion. 

A first  source of inspiration for answering these questions comes from ideas of geometric quantization, 
also at the roots of spin foam models \cite{BC}. Let us see this in the case of the Boulatov model (\ref{boulatov}): upon Peter-Weyl expansion into  irreducible $\SU(2)$ representations labelled by $j_i \!\in\! \frac12 \N$, the gauge invariant field is expressed in terms of three valent-invariant  tensors 
$$\imath^{j_1, j_2, j_3} \in \mbox{Inv}\left[\cH_{j_1}\otimes\cH_{j_2}\otimes \cH_{j_3} \right]$$ 
In this picture, the Boulatov field represents a quantum triangle with fixed lengths: 
each $\cH_{j_i}$ represents the Hilbert space of states of a quantized vector of length $j_i$,  upon geometric quantization that promotes 
the classical edge vector $\vec{x}_i$ into the $\SU(2)$ generators $\widehat{X}_i$; the classical closure  condition $\vec{x}_1 + \vec{x}_2 + \vec{x}_3 \!=\! 0$ for a triangle then projects into the subspace of invariant tensors. Geometric quantization here also gives us the glueing rules for triangles as traces of tensors. 
In such a representation of group field theory, the Feynman amplitudes take the form of spin foam models. 
The challenge has been to export these ideas to four dimensions for the construction of realistic quantum gravity models. 

An elegant, complementary  approach \cite{aristidedaniele}, which has the advantage of making the geometry content of GFT more manifest, has recently been  proved fruitful \cite{aristidedaniele23}. 
In this approach, the field variables are data encoding classical simplicial geometry  (edge vectors in 3d, area bivectors in 4d...), and their 'quantization'  results in a non-commutative star-product on the space of fields. 
This non-commutative structure rests on the phase space structure of the discrete classical theories these GFT models are set to quantize, usually given by the cotangent bundle of a group manifold, with configuration space being a Lie group and the dual space being the corresponding Lie algebra; it thus encodes the classical non-commutativity of the `momentum' space of the classical theory.
In the case of the model (\ref{boulatov}), upon so-called group Fourier transform \cite{FT}
\be
\vphihat(x_1,\cdots x_3)\!:=\! \int[\extd g_i]^3\, \varphi(g_1, \cdots g_3)\, \e^{i \Tr x_1 g_1} \cdots \e^{i \Tr x_3 g_3}
\ee
the model is expressed in terms of fields on copies of $\su(2)\sim \R^3$ endowed with a non-commutative star-product (dual to group convolution). 
Gauge invariance translates into a closure condition $x_1+x_2+x_3 \!=\! 0$ for the variables, thus naturally interpreted as the edge vectors of a (non-commutative) triangle. 
Glueing rules for the triangles are then simply dictated by the star product, in a way that identifies the edge vectors of the common edges by means of non-commutative $\star$-delta functions. This result in the following expression for the action  (\ref{boulatov}): 
\be \label{action in x}
S [\vphihat]  = \frac12 \int [d^3 x_i]^3 \, \vphihat_{123} \star \vphihat_{321} - \frac{\lambda }{4!} \, \int  [d^3 x_i]^6\, \vphihat_{123} \star \vphihat_{345}\star \vphihat_{526} \star \vphihat_{641}
\ee
where the star-product relates pairwise the variables $x_i$ with the same index $i$. 
In such a representation of group field theory, the Feynman amplitudes take the form of simplicial gravity path integrals. 
Remarkably, this approach easily extends to 4d gravity models.

\

(v) {\bf What are the symmetries of a GFT model and what are their consequences?}

\

\noindent We start entering a less explored territory, despite the great importance of the issue. Symmetries are in fact crucial, in a field theory context, to constrain model building and understanding the true nature of the models, to define appropriate observables, to identify universality classes of critical behaviour, to distinguish different phases of the system, and so on.  

Up to now, only a few symmetries of GFT models for topological and, even less, gravity 4d models are understood and under some control 
\cite{Boulatov, Ooguri, FloEtera, diffeo, joseph}.  
The GFT implementation of the symmetry which becomes, at the level of the GFT Feynman amplitudes, the local gauge invariance of simplicial path integrals is well-understood.  Recently, also the GFT symmetry corresponding to 3d simplicial diffeos (after gauge invariance is imposed) and to translation invariance in BF models has been uncovered \cite{diffeo}. Interestingly, all these transformations, which are local symmetries in the simplicial path integral, are global transformations of the GFT field, from a field-theoretic point of view. Moreover, diffeos/translations are implemented as {\sl quantum group} symmetries.  
These are clearly at least partially broken in 4d gravity models, but the details of this breaking have not been elucidated yet. 

Combinatorial symmetries with respect to permutations of the arguments of the GFT field have also been considered, but their implications beyond those on the structure of the Feynman diagrams have not been explored. A more careful study of GFT symmetries and of their consequences in GFT models, the 4d gravity ones in particular, is certainly needed. One class of transformations that would be interesting to identify is the GFT and tensor models analogue of the unitary matrix transformations that allow the re-writing of matrix models in terms of matrix eigenvalues (the true degrees of freedom). This re-writing is crucial for many of the key results in matrix models and one could expect a similar role in tensor models and GFTs.



\

(vi) {\bf Can one control the sum over diagrams and topologies?}

\

\noindent The perturbative sum is in fact pretty wild, including both simplicial manifolds of arbitrary topology and pseudo-manifolds  \cite{GFTgraphs}. 
A number of works have been devoted to a better understanding of the properties of the generated complexes. 
An important development was the definition of a GFT analogue of the large-N expansion of matrix models, dominated by a particular class of triangulations of the sphere, for simple tensor models and topological GFTs in any dimension \cite{largeN}. The GFT analogue of the size  $N$ of the tensors is a cut-off in the momenta (group representations). This expansion is yet to be understood in its physical significance and in its subdominant terms, but it opens the way for several applications, both in quantum gravity and statistical physics. Mathematically, it is complemented by a variety of results on the scaling behaviour with the cut-off of the amplitudes of topological BF models (and some more limited results on 4d gravity models), including rigorous power counting results \cite{scaling}. 
For some models, it was also possible to prove the Borel summability of the whole perturbative series \cite{Borelsum}. 

\

(vii) {\bf Is the GFT perturbative expansion physically useful?}

\

\noindent 
From the GFT point of view,  the discrete data appearing in the GFT action and amplitudes are physical, in the sense that they correspond to true degrees of freedom of the theory, and not mere discretization artifacts.  However, they are obviously very different from the variables we are used to adopt for describing continuum spacetime physics. From simplicial gravity, we know that these data correspond to (piecewise-flat) singular continuum geometries, which in turn can be used to reconstruct smooth geometries in the limit in which large numbers of the same data are considered (at least at the classical level). This points to the non-perturbative sector of the GFT formalism as the locus of continuum geometric physics (as indeed shown explicitly in the simple case of matrix models). However, being physical, even the data associated to finite (sums of) simplicial complexes (e.g. those obtained truncating the GFT perturbative sum to finite order) could in principle be used to describe physical phenomena, in the same way in which a finite number of particles could be used, in some limited circumstances, to mimic the behaviour of a fluid. What is hard, however, is to pinpoint what these circumstances are, exactly. In general, being simplicial gravity path integrals, one would say that the Feynman amplitudes of GFTs describe correctly gravitational phenomena to the extent in which piecewise-flat Regge geometries can be used to approximate them. This means we could take advantage of the expertise developed, in the use of such approximations, in numerical relativity. However, one could expect that the perturbative GFT expansion is more and more physically relevant, and convenient as a mathematical language, the  closer we are, physically, to the spacetime configuration represented by the GFT perturbative vacuum, which is unfortunately a very degenerate configuration in which no geometry and no space is actually present. 

\

(viii)  {\bf Why does one want to have a renormalizable GFT? is it even possible?}

\

\noindent Being a field theory, and aiming at a fundamental description of space and time down to the Planck scale, a GFT model should be renormalizable. Non-renormalizability would instead implies that the GFT model considered can at best be an effective description of a more fundamental theory. 
The renormalization group flow of a given GFT will tell us what type of interaction (thus, which dynamics) is relevant at various scales. 

The question becomes then whether one can hope to obtain such a renormalizable GFT \cite{gftrenorm}. The main source of doubt probably comes from the non-renormalizability of gravity around Minkowski spacetime. However GFTs are background independent formulations of gravity (hopefully), not limited to fluctuations of the metric around a given continuum spacetime geometry, and their perturbative expansion is formulated around a very different vacuum, one that corresponds to no-space at all. 
Being formulated in terms of fundamental structures that are either non-geometric in nature or anyway only associated to discrete geometries, it is not obvious what effective continuum degrees of freedom it captures. While this is obviously a difficulty in relating GFTs to continuum physics, it also warns us against the direct application of conventional wisdom about continuum perturbative quantum gravity. 

Studies of tensor models and GFT renormalization are just beginning \cite{josephvalentin}, but already one example of a fully renormalizable non-trivial tensor model in 4 dimensions, though much simpler than full-blown GFTs for 4d gravity, has been recently exhibited \cite{renmodel}. 

\

(ix) {\bf Where should one look for continuum gravity?}

\

\noindent The dynamics of smooth geometries should emerge as an effective dynamics for many degrees of freedom, i..e from large boundary graph-based states and the re-summation of infinite Feynman diagrams. In other words, continuum gravitational physics resides in the non-perturbative sector of the current definition of GFTs. This means looking for different vacua of the fundamental GFT and for the dynamics around them, possibly accompanied by `large-scale' and semi-classical approximations. Notice, however, that not only the semi-classical limit is a distinct limit from the continuum one (the latter being naturally a limit of increasing number of degrees of freedom, regardless of their quantum or classical nature), but the appropriate continuum limit could even emerge because of the quantum properties of the fundamental degrees of freedom, which should not be washed out beforehand. The example that comes to mind is that of quantum liquids. Obviously, this is {\it terra incognita}, at present, so it is hard to do more than speculations and some educated guesses.

If more and more interacting  degrees of freedom are to be taken into account on the way to the continuum limit, and if the theory has to be trusted over a wide range of scales (as a fundamental theory of quantum gravity should be), a phase transition should occur. 
Physically, the occurrence of such a phase transition switching from  microscopic excitations to collective ones, would correspond to a `geometrogenesis' \cite{graphity}, the true birth of continuum spacetime and geometry from a fundamental non-geometric phase. 
This scenario has been proposed repeatedly in the GFT context \cite{danieleEmergence, gftfluid, gftrenorm} and outside it \cite{seiberg}. 
We are gaining experience of phase transitions following from the resummation of the perturbative series in the large-N limit, in simple \cite{phase} and less simple \cite{phase2} tensor models. 
But more work is needed to establish such scenario in more involved GFT models, and to understand its physical significance. 

\

(x) {\bf Can we do physics with GFTs?}

\

\noindent In light of the above scenario, 
the answer is unfortunately  `not yet', given the current status of our understanding of the subject. The task is to improve our control over the non-perturbative sector of GFT and our tools for studying phase transitions in GFTs. 
In the meantime, the exploration of approximation methods for GFT has started, 
to study \cite{GFThydro,eterawinston,noi,EffHamilt} the effective dynamics for fluctuations around non-trivial GFT field configurations by expanding the  GFT action 
around them, interpreted as non-trivial quantum spacetime backgrounds.
The results (including the derivation of effective geometrodynamic equations, effective Hamiltonian constraint operators, emergent non-commutative matter field theories) are very interesting, but there is clearly still a long way  to go. In parallel, the construction of simplified GFT-like models has also started, in particular models with possible cosmological interpretation and application \cite{gfc}. The idea is that such simplified models could either be derived from or at least guide the mean field theory treatment of fundamental GFT models. Again, this is just a beginning.

\section{Conclusions}

\

We have answered, to the best of our current understanding of the subject, some questions about the GFT approach to quantum gravity, its relation with LQG and other approaches, and its present status. In the process, we have mentioned many of the important results obtained recently in this context, and outlined some direction for future work. We hope to have well represented the fast growth of this area of research, and convinced the reader of its promise. At the same time, we hope it stands clear that much more work is needed to address many of these questions fully, and in particular those whose answer will really establish GFTs as a solid candidate formalism for quantum gravity.

\newpage

{\bf References}

\


\end{document}